\documentstyle[aps]{revtex}
\begin{document}
\draft
\title{Pionic Bound States in Momentum Space}
\author{Dinghui H. Lu and Rubin H. Landau}
\address{Physics Department, Oregon State University, Corvallis, OR 97331}
\date{\today}

\maketitle

\begin{abstract}%--------------------------------------------------
Bound states of the pion--nucleus system are investigated in momentum
space using a microscopic optical potential with inherent energy
dependences and nonlocalities arising from elementary potential
models.  The wave equation and computational techniques are tested,
shallow levels are calculated, and a comparison is made with
experiment. Deep levels are found to be deeper and broader than those
in other investigations.
\end{abstract}

%\pacs{PACS numbers: 24.10-i, 24.10.Ht, 24.70.+5, 25.10.+5, 25.40.-h, 25.40.Cm}

\section{Introduction}

A complex potential is sometimes used in quantum mechanics to model a
physical system which is coupled to another system or channel. This
approach is used, for example, in analyzing exotic atoms where an
optical potential is added to the Coulomb potential to describe the
exotic particle's strong interaction and eventual annihilation with
the nucleus.  The normalizable states for such systems are
``quasibound'' since they decay exponentially in time with a
lifetime proportional to the inverse of the imaginary part of the
binding energy.

Because strong channel coupling is non perturbative, its phenomenology
is often non intuitive.  For example, if a potential has sufficiently
large imaginary (absorptive) part, an {\em increase} in the absorptive
strength may lead to a {\em decrease} in the level's width\cite{absp}.
Here the increase in the imaginary potential leads to further
absorption of the wavefunction from the nuclear interior, and the
corresponding decrease in the wavefunction's overlap with the nucleus
leads to a decrease in the level's width. In these cases, even if the
real part of the optical potential is attractive, the large imaginary
part acts as a repulsive potential by excluding the wavefunction from
the nucleus.

Recently, theoretical studies by Friedman and Scoff\cite{friedman},
Hirenzaki et al.\cite{hiren}, and Toki and
Yamazaki\cite{toki88,toki89} have pointed out that for deeply bound
levels in heavy nuclei, the pion optical potential might be
sufficiently absorptive for there to exist states of the type just
described.  Furthermore, these quasibound states would also be
``quasistable'' (and thus observable) since their widths would be
smaller than the surrounding energy--level spacings ($\Gamma \simeq
\Delta E/10$). For example, $\mbox{}^{208}$Pb would have $1s$ and $2p$
levels bound by $\sim 7$ and $\sim 5$ MeV and widths $\simeq 0.5$ MeV
--- in contrast to the $\sim 20$ MeV widths expected if lowest--order
perturbation theory were valid.  Furthermore, although the pion is
close enough for it to get bound up with the nucleus, the absorption
so excludes the wavefunction that the states become hybrid
Coulomb--nuclear ones with the dominant binding arising from the
Coulomb potential, and the probability density peaking just outside
the nuclear surface.

Observing deeply bound pion states in heavy nuclei is difficult
because the X-ray cascade ends with the pion being absorbed before it
reaches the deep levels. Transitions to the $1s$ level are seen
experimentally only for $Z<12$ and to the $2p$ level only for $Z<30$,
and so alternative observational methods have been proposed.  A recent
experimental search using the $(n,p)$ transfer reaction did not record
the distinct peaks expected for a $1s$ or $2p$ transition in
Pb\cite{iwasaki}.  Apparently, the cross sections are smaller, the
widths larger, or the binding energies less certain than predicted.
Other suggested experiments include $(n,d)$ pickup\cite{toki91},
$(\pi^{-},\gamma)$ radiative trapping\cite{spain92}, and $(\pi^{-},p)$
pickup\cite{kaufmann}.

The calculations of deeply bound pion states generally use a
phenomenological, coordinate space, optical potential of the
Ericson-Ericson form:
\begin{equation} \label{ericson}
V^{opt}(r) \simeq \frac{-4\pi}{2 \mu} \left[ b_{0} \rho(r) + c_{0}
\nabla\cdot\rho(r)\nabla + B_{0}
\rho^{2}(r) + C_{0} \nabla\cdot\rho^{2}(r)\nabla \right]
\end{equation}
where for clarity we assume here an isoscalar nucleus and leave off
the Lorenz--Lorentz factor. The parameters $b_{0}, c_{0}, B_{0},$ and
$C_{0}$ generally are determined by fits to less--bound energy levels
in pionic atoms. Because the pion has light mass, relativistic effects
are expected to be important, and because the pion has zero spin, the
correct dynamical equation is assumed to be the Klein--Gordon
equation:
\begin{eqnarray} \label{exact}
\left[E-V^{coul}(r)\right]^{2} \psi(r)
&=&\left[ -\nabla^{2} + \left\{\mu + V^{opt}(r)\right\}^{2}\right] \psi(r)\\
\label{kge} \left[E-V^{coul}(r)\right]^{2}\psi(r)
&\simeq& \left[ -\nabla^{2} + \mu^{2} + 2 \mu V^{opt}(r) \right] \psi(r)
\end{eqnarray}
In (\ref{exact}) the optical potential is treated as a scalar and
added to the mass $\mu$ and the Coulomb potential is treated as the
time component of a four--vector and subtracted from the energy $E$.
Because the square of the optical potential introduces technical
difficulties, the approximation (\ref{kge}) with no $(V^{opt})^{2}$ is
usually used\cite{friedman,toki89,spain93}  (Kwon and
Tabakin\cite{tabakin} have indicated how the quadratic term can be
handled in momentum space).

The investigation we report upon here was stimulated by the question
of how much the properties of deeply bound pionic states would change
if a different, possibly more microscopic, form for the optical
potential and wave equation were used. In particular, we wondered
about using {\em LPOTT}, the potential we had previously developed for
scattering\cite{lpott,l+t}, but never used for bound states.  {\em
LPOTT} is a theoretical optical potential in momentum space having
nonlocality, off--shell unitarity, and energy dependence (in addition
to its momentum dependence) arising from that of a pi--nucleon $T$
matrix derived from a separable potential. Since states bound by $\sim
10$ MeV are quite far down in energy from those at which the constants
$b_{0}, c_{0}, B_{0},$ and $C_{0}$ were fit, we particularly wondered
if {\em LPOTT}'s inherent energy dependence might make a difference.
Other factors of interest and difference include finite $\pi N$
interaction range effects in both the first and second order
potentials, covariant kinematics and transformation of scattering
amplitudes (``angle transformation''), and the use of the relativistic
Schr\"{o}dinger equation.  (Since the relativistic Schr\"{o}dinger
equation is also used in the definition of the $\pi N$ $T$ matrix and
in the derivation of the optical potential, our framework is
consistent, which unfortunately, use of (\ref{ericson}) in (\ref{kge})
is not.)

In \S\ref{formal} we describe the equations and computational method
used. In \S\ref{rel} we examine relativistic effects and the
numerical precision of our computations, and in \S\ref{shifts} make
comparisons to the experimental energies and widths of shallow states.
Finally, in \S\ref{deep} we examine the deep states supported by our
potential, and in \S\ref{nuc} deduce what changes in the potential
would be needed for there to exist pure nuclear pion bound states.
This in turn leads to further understanding of the low energy optical
potential.

\section{Formalism and Computational method} \label{formal}

The momentum space relativistic Schr\"{o}dinger equation is
\begin{eqnarray}
H |\psi \rangle &=& E |\psi \rangle \label{eigen}\\
 K(k) \psi_{l}(k) + \frac{2}{\pi} \int_{0}^{\infty}
V_{l}(k,k^{'};E) \psi_{l}(k^{'}) {k^{'}}^{2} dk^{'} &=&E \psi_{l}(k)
\label{se} \\
 \sqrt{m_{\pi}^{2}+ k^{2}} +
\sqrt{m_{A}^{2}+k^{2}} -m_{\pi} - m_{A}  &=& K(k) \label{K}
\end{eqnarray}
Here $K(k)$ is the relativistic kinetic energy and $V_{l}(k,k^{'};E)$
is the energy--dependent pion-nucleus potential after partial wave
projection.  There is no difficulty in momentum space with the
square roots in $K(k)$ since they contain just numbers not operators.
Although the relativistic Schr\"{o}dinger equation
(\ref{se})-(\ref{K}) obviously contains relativistic kinematics, it is
not covariant. However, it is the direct part of the
Blankenbecler--Sugar equation, which is the 3D reduction of the
Bethe--Salpeter equation, a covariant, two--particle
equation\cite{coester,qmII}. The Klein--Gordon equation, in turn, is
relativistic and covariant, but it is only a one--particle equation,
and when used as in (\ref{kge}) does not treat the optical potential
covariantly.

To handle the most general case of coupled and open channels, we
transform (\ref{eigen}) to an integral equation involving the Green's
function (where the appropriate boundary conditions are
incorporated)\cite{landau83}:
\begin{eqnarray}
|\psi\rangle &=& G_{E}V_{E}|\psi\rangle
	\label{G}\\
\psi_{l}(k) &=& \frac{2}{\pi}
	\frac{1} {E- K(k)} \int_{0}^{\infty}
	V_{l}(k,k';E)\,\psi_{l}(k')  k'^{2} dk'
\end{eqnarray}
Note that for bound states there is no incident wave in (\ref{G}) and no
$+i\epsilon$ in $G_{E}$. If we consider $V_{E}, K$, and $G_{E}$ as
operators or their matrix representation, then we see that solution of
the eigenvalue problem (\ref{eigen}) is equivalent to demanding
nontrivial solutions of (\ref{G}), namely,
\begin{equation}
 \left(1 - G_{E}V_{E}\right)|\psi\rangle  = 0 \ \
\Rightarrow \ \ det |1 - G_E V_{E}| = 0 \label{det}
\end{equation}
We search for solutions of (\ref{det}) in complex energy space after
removing the Coulomb singularity with the Lande' subtraction
technique. Solving (\ref{det}) is equivalent to determining the
energies of the poles of the $\pi$-nucleus $T$ matrix\cite{landau83}.
In contrast, the momentum space formulations of Kwon and
Tabakin\cite{tabakin} and Ciepl\'{y} et al.\cite{mach} directly solve
(\ref{se}) as an eigenvalue problem (Kwon and Tabakin use the Lande'
subtraction and Ciepl\'{y} et al. use the Vincent--Phatak cutoff).  In
further contrast, Kalbermann et al.\cite{gal} sidetrack momentum space
entirely and include finite range effects (but not the full energy
dependence of a separable $\pi N$ T matrix), by solving an
integro--differential wave equation in coordinate space.

In our calculations, the potential $V_{E}$ is taken as the sum of
electromagnetic plus optical potentials:
\begin{eqnarray}
\langle\vec{k}^{'}|V_{E}|\vec{k}\rangle &=& V^{coul}(\vec{q}) +
V^{VP}(\vec{q}) + V^{OP}_{E}(\vec{k}^{'},\vec{k})\\
V^{coul}(\vec{q}) &=&
-\frac{Z \alpha \rho_{p}(q)}{2\pi^{2} q^{2}} \label{Vcoul}\\
V^{VP}(\vec{q}) &=& -\frac{Z\alpha^{2}\rho_{p}(q)}{2\pi^{3}}
\int_{1}^{\infty}   \frac{ (2t^{2}+1)(t^{2}-1)^{1/2}}
{3t^{4}(q^{2}+(2t/ \lambda_{e})^{2})} dt
\end{eqnarray}
Here the momentum transfer $\vec{q} = \vec{k'}-\vec{k}$, $V^{coul}$ is
the Coulomb potential for a nucleus of finite size, $\rho_{p}(q)$ is
the proton form factor, $V^{VP}$ is the order--$Z\alpha^{2}$ vacuum
polarization (Uehling) potential\cite{tabakin,vac}, and $\lambda_{e}$
is the electron Compton wavelength.  The optical potential $V^{OP}$ is
evaluated in the impulse and factored approximations, and contains
terms of first and second order in $\rho$:
\begin{eqnarray}
V^{OP}_{E}(\vec{k}^{'},\vec{k}) &=& U_{E}^{(1)}(\vec{k}^{'},\vec{k}) +
U^{(2)}_{E}(\vec{k}^{'},\vec{k}) 	\label{U}\\
U^{(1)}_{E}(\vec{k^{'}},\vec{k}) &=& Z \langle \vec{k^{'}},\vec{p}_{0}
-
\vec{q}\left|t^{\pi p}_{\omega_{3B}}\right|\vec{k},\vec{p}_{0}\rangle
\,\rho_p(q) \ + N \langle \vec{k^{'}},\vec{p}_{0} -
\vec{q}\left|t^{\pi p}_{\omega_{3B}}\right|\vec{k},\vec{p}_{0}\rangle
\,\rho_n(q)\ \label {U1}\\
	&=& \frac{\gamma} {2\pi^{2}}  \sum_{ilj}
\frac { g_{ilj}(\kappa^{'}) g_{ilj}(\kappa)}
 { D_{ilj}(\omega_{3B})}
P_{l}(\cos\theta_{\kappa^{'}\kappa})\rho_{i}(q)
\label{sep} \\
D_{ilj}(\omega) &=& \frac{1}{\sigma_{ilj}} -\frac{2}{\pi}
\int_{0}^{\infty}  \frac{g_{ilj}^{2}(\kappa)} {\omega
- E_{\pi}(\kappa)
- E_{N}(\kappa) + i\epsilon} \kappa^{2} d\kappa  \label{D} \\
U^{(2)}_{E}(\vec{k^{'}},\vec{k}) &=& -\frac{4\pi A^{2}}{2\mu_{\pi
A}(2\pi)^{3}} [ B_{0}(E) \frac {\hat{g}_{s}(\kappa)\hat{g}_{s}(\kappa^{'})}
{\hat{g}_{s}^{2}(\kappa_{0})} + C_{0}(E)
\frac {\hat{g}_{p}(\kappa)\hat{g}_{p}(\kappa^{'})}
 { \hat{g}_{p}^{2}(\kappa_{0})} \vec{\kappa^{'}}\cdot \vec{\kappa}]
\rho^{2}(q) \label{U2} \\
\hat{g}_{s}(\kappa) &=& \frac {1}
{\alpha_{s}^{2}+\kappa^{2}},
\ \ \hat{g}_{p}(\kappa) = \frac {1} {(\alpha_{p}^{2}+\kappa^{2})^{2}}
\end{eqnarray}
Here $ilj$ labels the $\pi N$ eigenchannel ($i=$ proton, neutron, or
isospin, $l=$ orbital, and $j=$ total angular momentum),
$\sigma_{ilj}$ is the sign of the potential in that eigenchannel, and
$\kappa$ and $\kappa'$ are the initial and final $\pi N$ {\sc COM}
momenta.  The nuclear form factors $\rho_{p,n}(q)$ in (\ref{U1}) and
(\ref{U2}) are Fourier transforms of Woods-Saxon
densities\cite{lpott}.  To aide comparison with other optical
potentials, we take the size parameters as the standard set used by
Friedman et al.\cite{eli} with extension to Pb by use of the charge
parameters from De Vries et al.\cite{rhos}. The parameters are given
in Table~\ref{tab.rho}.

For simplicity of calculation we have used the factored form of
$U^{(1)}$ (\ref{U1}), which clearly does not average over the
nucleons' Fermi momenta. We can remove this limitation, yet the
computation would be longer, more complicated, and possibly less
reliable as we extrapolate deep down in energy. We do include nucleon
recoil in part, by making the optimal choice for the momentum of the
struck nucleon\cite{l+t},
\begin{equation}
\vec{p}_{0} = -\frac{\vec{k}}{A} + \frac{(A-1)\vec{q}}{2A}
\end{equation}
This $\vec{p}_{0}$ is optimal in producing the best factored
approximation and the most physical off--energy--shell kinematics. We
include further recoil and Fermi motion effects in our use of the
three--body subenergy $\omega_{3B}$ as the $\pi N$ energy in
(\ref{U1}). This choice arises from a three--body model for the
optical potential\cite{l+t} in which there is a pion of momentum
$\vec{k}$, a nucleon of momentum $\vec{p_{0}}$, and a core of momentum
$\vec{P}=-(\vec{k} + \vec{p} +\vec{p_{0}})$. The $\pi N$ $T$ matrix is
then evaluated at an energy which is the magnitude of the difference
in energies of the $\pi N$ pair and the core:
\begin{equation}
\omega_{3B}^{2}(E) = (k_{\pi}^{\mu} + k_{A}^{\mu} -P^{\mu})^{2}
 \simeq \left[ E + m_{\pi} + m_{A} - E_{A-1}(P) - E_{B}\right]^{2} -P^{2}
\end{equation}
Here $E_{B}$ is an effective core--nucleon binding energy we take as
$20$ MeV. Changing $E_{B}$ to $10$ MeV would increase our final
$\Gamma_{1s}$(Pb) by $\sim 20 \%$, and $E_{re}$(Pb) by $\sim 1\%$.

Pauli blocking is included by modifying the $\pi N$ T matrix for a
nucleon embedded in a Fermi sea with Fermi momentum $P_{F}$ appropriate to the
nucleus in question\cite{l+t}. Specifically we modify the $D$
functions in (\ref{D}) with an  angle--averaged Pauli operator $Q_{0}$:
\begin{eqnarray}
D_{\alpha}(\omega) \rightarrow D_{\alpha}^{(P)}(\omega) &=&
\frac{1}{\sigma_{\alpha}} -\frac{2}{\pi}
\int_{\xi P-P_{F}}^{\xi P+P_{F}} \frac{g_{\alpha}^{2}(\kappa)
	Q_{0}(P,\kappa)}
{\omega - E_{\pi}(\kappa) - E_{N}(\kappa) + i\epsilon }
\kappa^{2} d\kappa \label{DP} \\*[1ex]
Q_{0}(P,\kappa) &=& \left\{
	\protect{\begin{array} {ll} 0, &
\xi P + \kappa < P_{F}\\ 1, & |\xi P - \kappa|> P_{F}\\
\overline{Q_{0}}, & \mbox{otherwise} \end{array}}
\right. \\*[1ex]
\overline{Q_{0}} &=& \frac{ (\xi P + \kappa)^{2} - P_{F}^{2} } {4\xi P
\kappa}, \ \ \ \xi = \frac{m_{N}}{m_{N}+m_{\pi}}
\end{eqnarray}

The two-body $T$ matrices and momenta in (\ref{U1}) are related to
off--shell ones in the $\pi N$ {\sc COM} by a Lorentz covariant
prescription\cite{l+t} which determines $\gamma$, $\vec{\kappa}$, and
$\vec{\kappa^{'}}$. As indicated by the $g_{ilj}$ product in
(\ref{U1}), we include the finite range of the $\pi N$ interaction by
using a separable potential\cite{lpott} which, in turn, was fit to
$\pi N$ scattering data.  Note, the explicit energy dependence given
by the solution of the $\pi N$ T matrix in (\ref{U1})-(\ref{sep}) is
important for the present study of deeply bound states because it
permits a physically motivated evaluation of optical potential far
below the $\pi N$ threshold. Since our model has finite range $\pi N$
interactions, inclusion of a Lorentz--Lorenz term is
inappropriate\cite{l+t,gal}.

The second order potential $U^{(2)}$ includes contributions from
processes such as virtual nuclear excitation and pion annihilation
(true absorption). It is particularly important for pionic atoms since
its imaginary part provides absorption. The form we use for $U^{(2)}$
in (\ref{U2}) is a generalization of the separable form used for
$U^{(1)}$ (we take $\kappa_{0} \rightarrow 0$ for subthreshold
extrapolation).  Unfortunately, while tremendous theoretical and
experimental progress has been made in understanding the contribution
of true pion absorption to the pi-nucleus interaction, that progress
is not far enough along to permit a determination of $U^{(2)}$ with
enough accuracy for present purposes. Consequently, the quantities
$B_{0}$ and $C_{0}$ are fit to the $1s$ and $2p$ levels' shifts and
widths in pionic $^{16}$O and $^{40}$Ca. In addition, while for
scattering we modeled the energy dependence of $B_{0}$ and $C_{0}$ as
that of the $\pi^{+} D \rightarrow pp$ cross section, we have not been
able to determine a reliable, analytic continuation for subthreshold
energies using the energy dependence of either $\sigma(\pi^{+} D
\rightarrow pp)$ or the parameters of Ciepl\'{y} et al.\cite{mach}.
While our estimates indicate that the subthreshold energy dependence
is weak, our ignorance of it weakens our model.

Since our solution of (\ref{det}) determines the complex eigenenergies
but not eigenfunctions, once the eigenenergies are known we solve for
the momentum space wavefunctions by using inverse
iteration\cite{tabakin} on the Schr\"{o}dinger equation (\ref{se}):
\begin{eqnarray}
	\left[H_{E}\right] \left[\psi_{l}(k_{i})\right]
&=& E \left[\psi_{l}(k_{i})\right] \\
\left[\psi_{l}(k_{i})\right]
&=& \left[H_{E}-E\right]^{-1} \left[\psi_{l}(k_{i})\right]
\label{psik}
\end{eqnarray}
The wavefunction normalization  is then
\begin{eqnarray}
	1 &=& \int_{0}^{\infty} \,\psi^{2}_{l}(k) \,k^{2}dk
\label{norm}\\
\Rightarrow \ \ 1 &=& \int_{0}^{\infty} \,\left[(\mbox{Re~}\psi_{l}(k))^{2}
- (\mbox{Im~}\psi_{l}(k))^{2}\right]\,k^{2}dk
\label{renorm}\\
\label{imnorm}
0 &=& \int_{0}^{\infty} \,\mbox{Re~}\psi_{l}(k) \,\mbox{Im~}\psi_{l}(k)
\,k^{2}dk
\end{eqnarray}
Note that the normalization condition (\ref{norm}), which follows from
the Gamow state formalism of Hern\`{a}ndez and Mondragon\cite{wfp},
normalizes the square of the complex wave function rather than the
square of the modulus.  This in turn requires the overlap of
$\mbox{Re~}\psi_{l}(k)$ and $\mbox{Im~}\psi_{l}(k)$ to vanish
(\ref{imnorm}), which often introduces an oscillation in
$\mbox{Re~}\psi_{l}(r)$.  Once the momentum--space wavefunction is
known, it is transformed to coordinate space via:
\begin{equation}
	\psi_{l}(r) \equiv \frac{u_{l}(r)}{r} =
i^{l}\left(\frac{2}{\pi}\right)^{1/2} 	\int_{0}^{\infty}
\psi_{l}(k)\, j_{l}(kr) \,k^{2}dk,\label{psir}
\end{equation}

\section{Results and Discussions}

\subsection{Relativistic Effects and Numerical Precision}\label{rel}

Relativistic effects should be important for deeply bound states in
heavy nuclei. While our dynamical equation is the relativistic
Schr\"{o}dinger equation ({\ref{se}), in contrast to the more
conventional approximate Klein--Gordon equation (\ref{kge}), it is
reasonable to wonder if either equation is adequate for calculating
such states.  We examine the point Coulomb case since then we know the
analytic solutions to the Schr\"{o}dinger equation (the Bohr energies)
and to the Klein--Gordon equation\cite{qmII,bethe}:
\begin{eqnarray}
E_{nl}^{KGE} &=& \frac{ mc^{2} } { \{1 + (Z\alpha)^{2}/
	[n-l-\frac{1}{2} + \sqrt{ (l+\frac{1}{2})^{2} -(Z\alpha)^{2}}
 ]^{2}\}^{1/2} } \label{Ekge}\\*[1ex]
 	&\simeq& mc^{2} - \frac{mc^{2}(Z\alpha)^2} {2n^{2} }
	- \frac{mc^{2}(Z\alpha)^{4}} {2n^{4}}
\left(\frac{n}{l+\frac{1}{2}}
	- \frac{3}{4}\right) + O(\alpha^{6}) \label{expand}
\end{eqnarray}
To compare with the two--particle Schr\"{o}dinger theory, we use a
reduced mass $\mu$ in the second term of (\ref{expand}) and the pion
mass $m_{\pi}$ in the first term.  We examine in Tables~\ref{tab.rel1}
and \ref{tab.rel2} both $\pi-\mbox{}^{40}$Ca and $\pi-\mbox{}^{208}$Pb
because $Z\alpha$ is greater than $\frac{1}{2}$ for Pb which means the
Klein--Gordon, point--Coulomb solution (\ref{Ekge}) is pathological
for $l=0$ (the finite--nucleus case is not).

Columns 2 and 3 in Table~\ref{tab.rel1} indicate our numerical
precision by comparing the computed and analytic results for the
nonrelativistic Schr\"{o}dinger equation (we know of no analytic
results for the relativistic Schr\"{o}dinger equation). We see that
the numerical results are good out to the fifth decimal place, and
that the computed $2s$ and $2p$ levels are degenerate (as well they
should be) within numerical precision---even though the partial--wave
potentials are quite different. Columns 3 and 4 in
Table~\ref{tab.rel1} and columns 2 and 3 in Table~\ref{tab.rel2}
indicate that relativistic effects are significant: $2$\% for
$\mbox{}^{40}$Ca-$1s$, $70$\% for $\mbox{}^{208}$Pb-$1s$, and $48$\%
for $\mbox{}^{208}$Pb-$2s$. In contrast, columns 5 and 6 in
Table~\ref{tab.rel1} and columns 3 and 4 in Table~\ref{tab.rel2}
indicate that both the relativistic Schr\"{o}dinger equation and the
Klein--Gordon equation using the reduced mass, remove the degeneracy
with $l$ and include similar enough relativistic corrections for the
differences to lie in the fourth significant figure.  (The effect
would be bigger for $s$ states in Pb --- if only they existed for the
Klein--Gordon equation.)

Because these comparisons indicate that the inclusion of terms of
higher order in the potential are important, and because the deeply
bound states are hybrid nuclear-Coulomb, we conclude that ignoring the
$(V^{opt})^{2}$ term in the Klein--Gordon equation ({\ref{kge}) may
have a significant (probably several percent) effects upon the
answers. This is consistent with the smaller effect found by Kwon and
Tabakin\cite{tabakin} for the lower energy, and less relativistic,
$3d-2p$ transition in $K^{-}\,\mbox{}^{32}$S.

\subsection{Experimental Shifts of Shallow States} \label{shifts}

The only adjustable parameters in our theory are the annihilation
strengths $B_{0}$ and $C_{0}$. Because our ultimate interest is the
$1s$ and $2p$ levels (and because we want to avoid anomalies
associated with the $3d$ levels\cite{seki}), we fit the experimental
energies and widths for the $1s$ and $2p$ levels in pionic
$\mbox{}^{16}$O\cite{1sO16,2pO16} and $1s$ level in
$\mbox{}^{40}$Ca\cite{ca40-1,ca40-2}, and determined
\begin{equation}
	B_{0}=(-0.074+0.067 i)m_{\pi}^{-4}, \ \ \
	C_{0} = (0.051+0.069 i) m_{\pi}^{-6} \label{fit}
\end{equation}
Since $U^{(2)}$ provides the absorption in our model, the imaginary
parts of $B_{0}$ and $C_{0}$ should have the most similarity to the
work of others. The value for $\mbox{Im~} C_{0}$ in (\ref{fit}) is
similar to those found in the coordinate space models, while the value
for $B_{0}$ is larger, although the momentum--space pionic atom
calculation of Ciepl\'{y} et al.\cite{mach} are more relevant. They
fit the $1s$ levels of $\mbox{}^{12}$C and $\mbox{}^{16}$O, and the
$2p$ levels of $\mbox{}^{32}$S and $\mbox{}^{40}$Ca, and find $B_{0} =
(-0.093 + i 0.042)m_{\pi}^{-4}, C_{0} = (-0.125+0.090 i)
m_{\pi}^{-6}$. The differences probably reflect the differences in
$U^{(1)}$ and in the data fit.

We do not expect a theoretical potential to give the same
level of agreement as a phenonenological potential whose many
parameters have been determined in a global search. Nevertheless, in
Table~\ref{tab.atoms} and Figure~\ref{fig.atoms} we show a comparison
between the strong interaction shifts predicted by our potential (the
$+$'s) and various pionic atom
data\cite{1sO16,2pO16,ca40-1,ca40-2,ca44,bi}. We see very good
agreement for the heavier nuclei, good agreement for the lighter
nuclei, and an incongruently large deviation for $\mbox{}^{18}$O
(recall, we fit $B_{0}$ and $C_{0}$ to $\mbox{}^{16}$O and
$\mbox{}^{40}$Ca). In general, our level of agreement is comparable to
the finite range, momentum space calculations of Ciepl\'{y} et
al.\cite{mach} --- even though the potentials and calculational
framework differ.

\subsection{Deep States}\label{deep}

In Table~\ref{tab.deep} and Figure~\ref{fig.deepE} we show the
energies and widths of the deeply bound $\pi^{-}\mbox{}^{208}$Pb
states. The subscript $EM$ denotes the inclusion of only
electromagnetic interactions (Coulomb + vacuum polarization), $tot$
the combined $EM$ plus optical potentials, and $r1$, $r2$, and $r3$
the r-space calculations of Toki et al.\cite{toki89}, Nieves et
al.\cite{spain93}, and of Konijn et al.\cite{konijn}.  We see that our
calculated binding energies are slightly larger ($\sim 7\%$) than
those of Toki et al., Nieves et al., and Konijn et al.\cite{konijn},
and our widths are significantly larger: for the $s$ states, $\sim
20\%$ relative to Toki et al.\cite{toki89} and $\sim 200\%$ relative
to Nieves et al.\cite{spain93}; for the $p$ and $d$ states, $\sim
50\%$ relative to Toki et al. and $\sim 200-300\%$ relative to Nieves
et al..

Because the differences increase with binding, the cause
appears to be the energy dependences of the potentials, the different
wave equations used, and the different extrapolations in the complex
energy plane.  Since the differences are much greater for $\Gamma$
than $\mbox{Re~} E$, the different annihilation potential has much to
do with the differences (we use a finite range model, the others do
not).  We clearly are finding a much greater optical model dependence
than $(0.4, 15)\%$ in $(\mbox{Re~} E, \Gamma)$ found by Nieves et al.
(possibly their $7-75 \%$ phenomenological renormalization of
strengths reduced the model dependence they found).  In our
calculations, the vacuum polarization increases the binding by $0.5\%$
and the width by $2\%$, which may be significant, but not compared to
the model dependence.

To further understand the physics of the nonlocal optical potential
and test the momentum space calculation, we have studied the wave
functions of the deeply bound states. The momentum space wavefunctions
are calculated via (\ref{psik}) and the coordinate space ones via the
Bessel transform (\ref{psir}). In the top of Figure~\ref{fig.prwf} we
display the squared modulus of the momentum space wavefunction on a
logarithmic scale, and in the bottom of the figure we display the
coordinate space wavefunction $\psi_{l}(r)$, both for the $1s$ state
in Pb. We see in the comparison to the pure Coulomb wavefunction, that
the optical potential introduces structures into the momentum space
wavefunction similar to those present in the potential. These
wavefunctions are similar in both shape and magnitude to those of Toki
et al.\cite{toki89}, and should yield similar predictions when used in
a DWIA calculation of the formation rate of the state.

Examination of the $r$-space wavefunction in the bottom of
Figure~\ref{fig.prwf}, shows that the strong interaction repels the
wavefunction out from the nuclear interior and that there is a node in
$\mbox{Re~}\psi(r)$ even though this is a $1s$ state. This node is a
consequence of the existence of $\mbox{Im~}\psi_{l}(r)$ and the
complex normalization constraint (\ref{norm})-(\ref{imnorm}). The
repulsion is also evident in Figure~\ref{fig.prob} where we plot the
probability density\cite{prob} $|u_{l}(r)|^{2} \equiv
|r\psi_{l}(r)|^{2}$ when there is only the EM potential (dashed
curves) and when there is the EM plus nuclear potentials (solid
curves). (The node is now hidden by the $r^{2}$ multiplication.) We
see in this figure the repulsion for the $1s, 2p$ and $2s$ states in
Pb, while for $3d$ and higher states there is an attraction.  Similar
results were found by Toki et al.\cite{toki89}.

While some repulsion (and the narrow widths) arise from the strong
absorption of the wavefunction near the origin, the $p$ wave repulsion
occurs only for heavier nuclei.  To understand this, we go back to the
optical potential (\ref{U1}), and note that for nuclei which have $Z
\simeq N$, the $Zt^{\pi p} + Nt^{\pi n}$ term in the theoretical
optical potential is attractive in $p$ wave --- as confirmed by the
$p$ level shifts in Table~\ref{tab.atoms}.  Yet for nuclei which have
$N >> Z$ like Pb, the $t^{\pi n}$ term dominates and the $p$ wave
potential becomes repulsive. In the left part of Figure~\ref{fig.pots}
we present $V(p,p)$, the diagonal parts of the momentum space
potentials for the $1s$ state in Pb and the $2p$ state in
$\mbox{}^{40}$Ca. The potential in this figure is clearly repulsive
for the $1s$ state in Pb and attractive for the $2p$ state in Ca
(although not shown, the $2p$ potential in Pb is repulsive as
expected).  We also see in both cases that the imaginary part of the
potential is absorptive.

\subsection{Nuclear Bound States} \label{nuc}

We have found in our numerical searches that the optical potential is
attractive enough to produce broad pion--nucleus resonances, but not
attractive enough to produce pure nuclear bound states.  We wondered
how much stronger the optical potential is needed to support nuclear
states, and so undertook a series of computations in which we
progressively increased the strength of the optical potential while
keeping the Coulomb potential constant.  We found that the hybrid
Coulomb--nuclear states eventually became nuclear ones,  explicitly, for
the range of strengths
\begin{equation}
V^{coul} +  V^{opt}  \leq V  \leq   V^{coul} +  15\, V^{opt}
\label{scale}
\end{equation}
the complex binding energies found are:
\begin{eqnarray}
-(0.4+0.001i)\,\mbox{MeV} & \leq E_{Ca} \leq & -(5 + 5 i)\, \mbox{MeV}
\label{E1} \\
-(7+0.4i) \,\mbox{MeV} & \leq E_{Pb}  \leq & -(22 + 13i) \, \mbox{MeV}
\label{E2}
\end{eqnarray}

In Figure~\ref{fig.deepwf} we present the probability
density\cite{prob} $|u_{l}(r)|^{2}$ for the $1s$ state in $\pi$-Pb for
varying optical potential strengths. We see the hybrid state slowly
moving closer to the nucleus, until a transition takes place and at
fifteen--fold strength the state is localized completely within the
nucleus. When the optical potential is this strong, the state
remain bound, for both Pb and Ca, even if the Coulomb potential is
switched off.

The increase in binding and eventual nuclear state is expected for $p$
states in Ca where the level shifts in Table \ref{tab.atoms} are
attractive. We were surprised, however, to find it for $s$ states
where the level shift is repulsive. At least for local potentials,
increasing the strength of the potential does not change its effect
from repulsive to attractive.  If we go back and look in the left part
of Figure~\ref{fig.pots}, we see that $\mbox{Im~} V_{OP}(p,p)$ always has the
sign expected for an absorptive potential and that $\mbox{Re~} V_{OP}(p,p)$
appears attractive for the $2p$ state in Ca and repulsive for the $1s$
state in Pb. So this does not explain how we can get a $1s$ nuclear
state.

In an attempt to unravel the physics behind our nonlocal potential's
$1s$ bound states, we made use of our knowledge of the radial
wavefunction $u_{l}(r) \equiv r\psi_{l}(r)$ obtained via (\ref{psik})
and (\ref{psir}), to deduce an equivalent and local coordinate space
potential.  We assumed there is a $V_{equiv}(r)$ which when used in a
nonrelativistic Schr\"{o}dinger equation produces this same
$u_{l}(r)$:
\begin{equation}
V_{equiv}(r) = \frac{\hbar^{2}}{u_{l}(r)2\mu}
	\left[\frac{d^{2}u_{l}(r)}{dr^{2}}
+ \left(\frac{2\mu E}{\hbar^{2}}
- \frac{ l(l+1)} {r^{2}}\right) u_{l}(r) \right]
\label{veff}
\end{equation}
We evaluate the second derivative in (\ref{veff}) numerically using
the central difference algorithm, and obtain the equivalent potential
shown in the right hand side of Figure~\ref{fig.pots}. While we do not
expect such a $V_{equiv}(r)$ to be the most physical of potentials, it
is revealing, and the fact that the imaginary part of $V_{equiv}(r)$
is always absorptive lends some confidence.

We see in the lower right--hand corner of Figure~\ref{fig.pots} that the real
part of the $V_{equiv}(r)$ for the normal strength $2p$ state in Ca
has a dominant, long range attraction --- but also a short range
repulsion. And since $u_{l}(r)$ has no nodes, there is no division by
zero in (\ref{veff}) to cause a sign change in $V_{equiv}(r)$.  Yet we
also find that the character of the potential changes dramatically as
the nuclear bound state forms, with the potential acquiring an
oscillation in the surface --- much like the one expected from the
$\nabla \cdot \rho(r) \nabla$ term in the Ericson--Ericson potential.
We find the bound state getting more bound, the wavefunction
developing another node, and this  showing up in the effective
potential.

Much to our surprise, in the upper right--hand corner of
Figure~\ref{fig.pots} we see that $V_{equiv}(r)$ for the $1s$ state in
Pb has a central attraction in addition to a dominant, mid--range
repulsion.  This is true even at normal potential strength where the
changeover in sign occurs close to the $r$ value at which $\mbox{Re~}
\psi_{l}(r)$ has a sign change in Figure~\ref{fig.prwf}.  Accordingly,
we now believe that the low energy pion optical potential has an $s$
state repulsion on the outside of the nucleus and an attraction
within.  Furthermore, this shows how, as the nonlocal optical
potential is made stronger, the effective attractive part increases in
depth and eventually leads to a confined nuclear state.

\section{Conclusions}

We have shown again\cite{tabakin,mach} that pionic atom energies and
wavefunctions can be calculated accurately in momentum space with a
microscopic and nonlocal optical potential. Our particular potential
includes effects from finite $\pi N$ ranges in both the first and
second order terms, covariant kinematics, Pauli exclusion, nucleon
recoil, off--energy shell dynamics, and vacuum polarization.  The
Coulomb force is included exactly, and in our approach the energies
and widths are determined by searching for the poles of the $T$ matrix
for the combined Coulomb plus nuclear potential. With present
techniques we reproduce analytic results to five significant figures.
Although computations in momentum space use more computer time than
those in coordinate space, they can handle exactly various types of
nonlocalities and can use theoretical results with fewer
approximations; we also find them more elegant.  Given modern
computing techniques and machines, widespread momentum space
applications are now practical as shown by our brief survey of levels
throughout the periodic chart.

The focus of our calculations were the deeply bound ($1s$, $2s$, and
$2p$) levels in lead which appear as hybrid Coulomb-nuclear states.
Our calculations show these levels to be more bound and wider than
reported previously\cite{friedman,toki89,spain92}: $\sim 7\%$ deeper,
$\sim 20-200\%$ wider $s$ states, and $\sim 50-300\%$ wider $p$ and
$d$ states. The wavefunctions appear similar to those of Toki et
al.\cite{toki89} so we do not expect major changes in atomic formation
rates. Most importantly, the states remain quasi-stable
(nonoverlapping) and therefore are still experimentally observable.
There may, however, be some changes needed in the experimental search
for them.

The model dependence we have found appears to arise from our inclusion
of energy dependences in the optical potential, complex energy
extrapolations, and particularly our treatment of the finite--range
annihilation term. The use of a relativistic wave equation is
important for these states, and our use of the optical potential in a
relativistic Schr\"{o}dinger equation---as opposed to the approximate
Klein-Gordon equation which ignores the quadratic $V_{opt}^{2}$
term---may make some differences for these deeply bound states.

We view the major uncertainty in this work to be the description of
pion annihilation. We use a quasi--deuteron absorption model, but it
is not capable of predicting precise values for the annihilation
strengths $B_{0}$ and $C_{0}$ at threshold, or their energy
dependences below threshold.  In addition, there are unknown isovector
strengths to include for a nucleus such as lead, and they also appear
to be accessible only phenomenologically and at threshold.  A further
concern than just the values of the strengths, is the simple
assumption that the annihilation potential has $\rho^{2}$ dependence;
annihilation is a complicated process in which various numbers of
nucleons enter, and the $\rho^{2}$ dependence is only approximately
true if there were only two.

We have also looked for pure nuclear bound states of pions at much
deeper energies. We did not find any at normal densities and had to
increase the strength of the nuclear potential at least eight--fold
before any appeared. In the process of looking for these states we
discovered that the real part of the effective local potential for a
$1s$ pion in Pb, while predominantly repulsive, has an inner
attractive part. It is this attractive part which binds the pion
within the nucleus as the potential strength is increased. Observing
this dual character of the low energy potential experimentally would
be fascinating, but difficult; measuring pion scattering from nuclei
as a function of energy would not help much since the $\pi N$
interaction itself is highly energy dependent.  Likewise, the
attractive $p$ wave potential in Ca appears to have an inner part
which is repulsive, and this too appears to be a new finding.

\acknowledgements%------------------------------------

It is our pleasure to thank Eli Friedman for the conversations and
assurances which stimulated the present study. We also wish to thank
Iraj Afnan, Bill Kaufman, Anne Trudel, D. Frekers, Al Stetz, Carmen
Garcia-Recio, and Avraham Gal for illuminating discussions.  We
gratefully acknowledge support from the U.S. Department of Energy
under Grant \ DE-FG06-86ER40283  and the people at the National
Institute for Nuclear Theory, Seattle for their hospitality during
part of this work.

%-----------REFERENCES-----------------------
%\begin{thebibliography}{99}

%\end{thebibliography}

%Figure captions%=========================================

\begin{figure}
\caption{Comparison of optical potential predictions and
measurements of shifts (top) and widths (bottom) for the indicated
shallow levels of pionic atoms. Data are from references
\protect\cite{1sO16}-\protect\cite{bi}.}
\label{fig.atoms} %atomic shifts and widths
\end{figure}

\begin{figure}
\caption{Energy levels (top) and widths (bottom) of various deeply bound
$\pi^{-}\mbox{}^{208}$Pb states. The solid curves derive from the
present p--space potential, the others from the r--space potentials of
Toki et al.\protect\cite{toki89} and Nieves et
al.\protect\cite{spain93}. }
\label{fig.deepE} %deep levels
\end{figure}

\begin{figure}
\caption{Momentum space wavefunction $\psi_{l}(p)$ and coordinate space
wavefunctions $\psi_{l}(r)$ for $\pi^{-}Pb$ $1s$ state.
Dashed curves derive from the EM potential alone, solid curves
also contain the optical potential, and  the dot--dashed curve shows
$\mbox{Im~}
\psi_{l}(r)$ for the total potential.
A logarithmic scale is used for $\psi_{l}(p)$ to show details of
nuclear effect. In lower part of figure the exclusion of wavefunction
from the nucleus of radius is $\sim 7.1\,fm$ is evident as is the node
introduced into the wavefunction.}
\label{fig.prwf} %r space wf
\end{figure}

\begin{figure}
\caption{Probablity density $|u_{l}(r)|^{2} \equiv |r\psi_{l}(r)|^{2}$
for the $1s$, $2s$, and $2p$ $\pi^{-}Pb$ states.  Dashed curves
contain only the EM potential, solid curves EM plus optical
potentials. The exclusion of wavefunction from the nucleus of radius
is $\sim 7.1\,fm$ is evident, although the node in the real part of
the wavefunction at $\sim 2.5$ evident in Figure~\protect\ref{fig.prwf}
is not.}
\label{fig.prob} %r space wf
\end{figure}

\begin{figure}
\caption{The diagonal momentum space potentials $V(p,p)$ (left) and
equivalent  local coordinate space potentials $V_{equiv}(r)$ (right)
for the $1s$ state in $\mbox{}^{208}$Pb (top) and the $2p$ state in
$\mbox{}^{40}$Ca (bottom).}
\label{fig.pots} % potentials
\end{figure}

\begin{figure}
\caption{Probablity density $|u_{l}(r)|^{2} \equiv |r\psi_{l}(r)|^{2}$
of the $1s$ $\pi^{-}Pb$ wave function arising from an optical
potential increased, as in (\protect\ref{scale}), by 1 (solid), 8
(dot--dashed), and 15 (dashed) times its normal value. The 15--fold
increase leads to a nuclear bound state within the nucleus of radius
$\sim 7.1\,$fm.}
\label{fig.deepwf} %r space wf
\end{figure}

%\section{Tables}%--------------------------------------------
 %---------------------------
\begin{table}
\begin{center}
\caption{Parameters used in modified Wood-Saxon densities ($R$ and $a$ in
fermis).}
\label{tab.rho}
\begin{tabular}{rllllll}
     & $R_{p}$ & $a_{p}$ & $w_{p}$ & $R_{n}$ & $a_{n}$ & $w_{n}$ \\
\hline $^{16}O$ & 2.608 & 0.513 & -0.051 & 2.608 & 0.461 & -0.051 \\
$^{18}O$ & 2.634 & 0.513 & 0.0 & 2.72 & 0.447 & 0.0 \\ $^{40}Ca$&
3.669 & 0.584 & -0.102 & 3.669 & 0.584 & -0.102 \\ $^{44}Ca$&
3.70 & 0.55 & 0.0 & 3.82 & 0.505 & 0.0 \\ $^{108}Ag$& 5.32 & 0.52 &
0.0 & 5.47 & 0.473 & 0.0 \\ $^{208}Pb$ & 6.624 & 0.549 & 0.0 & 6.624
& 0.549 & 0.0 \\ $^{209}Bi$ & 6.609 & 0.545 & 0.0 & 6.88 & 0.5 & 0.0
\\
\hline
\end{tabular}
\end{center}
\end{table}

      %---------------------------
\begin{table}
\begin{center}
\caption{Effect of different treatments of relativity on point Coulomb
binding energies for $\pi^{-}\,\mbox{}^{40}$Ca atom. Compared are the
analytic Schr\"{o}dinger equation, the numeric Schr\"{o}dinger
equation, the relativistic Schr\"{o}dinger equation, the Klein--Gordon
equation with reduced mass, and the Klein--Gordon equation with pion
mass. All energies in KeV.}
\label{tab.rel1}
\begin{tabular}{llllll}
$nl$ & $E_{SE}(Anal) $ & $E_{SE}(Num)$ & $E_{RSE}$ &  $E_{KGE}(\mu)$&
$E_{KGE}(m_{\pi})$ \\  \hline
$1s$ & -1.4809 &   -1.4810 &  -1.5161  &  -1.5234  &  -1.5287   \\
\hline
$2s$ & -0.3702  &  -0.37025  &  -0.3761  & -0.3772  &  -0.3785    \\
$2p$ & -0.3702  &  -0.37024  &  -0.3714  &   -0.3717  &  -0.3730 \\
\end{tabular}
\end{center}
\end{table}

%---------------------------
\begin{table}
\begin{center}
\caption{Effect of different treatments of relativity on point Coulomb
binding energies (in MeV) for $\pi^{-}\,\mbox{}^{208}$Pb atom via
Schr\"{o}dinger equation, relativistic Schr\"{o}dinger
equation (numeric), and Klein--Gordon equation (which has no $s$ states for
$Z\alpha > \frac{1}{2}$).}
\label{tab.rel2}
\begin{tabular}{rrrr}
$nl$ & $E_{SE}$  & $E_{RSE}$ & $E_{KGE}$  \\  \hline
$1s$ & -24.97  &  	-42.53   & 	n \\
\hline
$2s$ & -6.243  &  	-9.224   &  	n \\
$2p$ & -6.242  &  	-6.584   &     -6.607  \\
\hline
$3s$ & -2.775  &  	-3.659  &   	n \\
$3p$ & -2.775  &  	-2.921  &     -2.929 \\
$3d$ & -2.774   & 	-2.824   &    -2.829  \\
\hline
$4s$ & -1.562   & 	-1.930  &  	n	  \\
$4p$ & -1.561   & 	-1.632   &  	-1.636    \\
$4d$ & -1.562   & 	-1.592   &   	-1.593    \\
$4f$ & -1.560   &  	-1.574  &    	-1.577 \\
\end{tabular}
\end{center}
\end{table}

%-----------------------------

\begin{table}
\begin{center}
\caption{Theoretical and experimental shifts and widths in KeV.}
\label{tab.atoms}
\begin{tabular}{llr|ll|ll}
  state& Ref   & $Re\,E$ & $\epsilon_{lpott}$ & $\epsilon_{exp}$
& $\Gamma_{lpott}$ & $\Gamma_{exp}$ \\ \hline
$1s\,\mbox{}^{16}$O &\cite{1sO16}  & -219.98 & -14.57 & -15.4 $\pm$ 0.1 & 8.14
&
7.92 $\pm$ 0.32  \\
$2p\,\mbox{}^{16}$O &\cite{2pO16}  & -59.06  & 13.4$\times 10^{-3}$ &
14.8  $\pm$ 1.6 $\times 10^{-3}$ & 5.5 & 6.8
$\pm$ 0.4 $\times 10^{-3}$ \\
$1s\,\mbox{}^{18}$O&\cite{1sO16}   & -215.76 & -18.87 & -19.9 $\pm$ 0.1 & 7.40
&
6.33 $\pm$ 0.43 \\
$2p\,\mbox{}^{40}\mbox{Ca}$ &\cite{ca40-1,ca40-2}   & -374.60 & 1.86  &
1.86  $\pm$ 0.08\tablenotemark[1] & 1.62  & 1.62  $\pm$ 0.11  \\
$2p\,\mbox{}^{44}$Ca & \cite{ca44}   & -374.78 & 1.92   & 1.58 $\pm$ 0.02  &
1.84  &
1.60$\pm$ 0.07   \\
$3d\,\mbox{}^{108}$Ag & \cite{ca40-2}  &-922.07 & 2.27  & 1.97 $\pm$ 0.03 &
1.24  &
1.41$\pm$  0.05   \\
$4f\,\mbox{}^{208}$Pb & \cite{ca44}   & -1582.19  & 2.28  & 1.67 $\pm$ 0.02 &
1.1  &
1.17$\pm$ 0.05\\
$4f\,\mbox{}^{209}$Bi & \cite{bi}   &-1622.21& 2.53  & 1.83 $\pm$ 0.06 & 1.31
&
1.24$\pm$ 0.14\\
\end{tabular}
\tablenotetext[1]{Average experimental values.}
\end{center}
\end{table}

%--------------------------------------------
\begin{table}
\begin{center}
\caption{Theoretical energies and widths in KeV for $\pi^{-} -
\mbox{}^{208}$Pb.  The subscripts $r1$, $r2$ and
$r3$ denote r--space calculations of
Toki et al.\protect\cite{toki89}, Nieves et al\protect\cite{spain93},
and Konijn et al.\protect\cite{konijn}.}
\label{tab.deep}
\begin{tabular}{rr|rrrr|rrrr}
$nl$ & $E_{EM}$ & $E_{lpott}$ & $E_{r1}$ & $E_{r2}$  & $E_{r3}$  &
$\Gamma_{lpott}$ & $\Gamma_{r1}$ & $\Gamma_{r2}$ & $\Gamma_{r3}$\\
\hline
$1s$ & -12302 & -7240   & -6959  & -6778  & -6924
  &   754  &    632   & 409 & 63 \\
$2s$ & -4492 & -3043  &  -2962 &  -2902 & -2954
&  208   & 183 &  140 & 13 \\
$2p$ & -5957 & -5399   & -5162 & -5105   & -5138
&  619  & 410   & 275 & 154\\
$3s$ & -2262 & -1682   & -1642   & -1613 & -1633
&   88  &  78   & 65 &5\\
$3p$ & -2707 & -2488   & -2418   & -2395  & -2408
&   201    & 151 &  99 & 52\\
$3d$ & -2831 & -2916  &  -2854 &   -2858  &
&   125    & 91 &   56 &\\
$4s$ & -1352 & -1067  &   -1045   & -1026 &
&   45    & 40 &   35& \\
$4p$ & -1538 & -1436 &  -1408  & -1394 & -1394
 &  90   & 71   &46 & 24\\
$4d$ & -1594 & -1639  &   -1606 &   -1606 &
&   73 &   52 &  31 &\\
$4f$ & -1579 & -1582  &   -1575 &   -1582 &
&   1.1    & 1.0    & 1.0& \\
\end{tabular}
\end{center}
\end{table}


\begin{references}

\bibitem{absp}
	R. Seki, Phys. Rev. C {\bf 5}, 1196 (1972); J.H. Koch, M.M.
Sternheim, and J.F. Walker, Phys. Rev. C {\bf 5}, 381 (1972); T.E.O.
Ericson and F. Schneck, Nucl .Phys. {\bf B19}, 450 (1970).

\bibitem{friedman}
	E. Friedman and G. Scoff, J. of Phys. G, {\bf 11}, (1985).

\bibitem{hiren}
	S. Hirenzaki, T. Kajino, K.-I. Kubo, H. Toki, and I. Tanihata,
Phys. Lett. B {\bf  194}, 20 (1987).

\bibitem{toki88}
	H. Toki and T. Yamazaki, Phys. Lett. B {\bf 213}, 129 (1988).

\bibitem{toki89}
	H. Toki, S. Hirenzaki, T. Yamzaki, and R.S. Hayano, Nucl.
Phys. A{\bf 501}, 653 (1989).


\bibitem{iwasaki}
	M. Iwasaki, A Trudel, A. Celler, O. K\"{a}usser, T.S.  Hayano,
R. Helmer, R. Henderson, S. Hirenzaki, K.P. Jackson, Y. Kuno, N.
Matsuoka, J. Mildenberger, C.A. Miller, H. Outa, H. Sakai, H. Toki, M.
Vetterli, Y. Watanabe, T. Yamazaki, and S. Yen, Phys. Rev. C {\bf 43},
1099 (1991).

\bibitem{toki91} H. Toki, S. Hirenzaki and T. Yamazaki,
	Nucl. Phys. {\bf A530},  679 (1991).

\bibitem{spain92}
	J. Nieves and E. Oset, Phys. Lett. B{\bf 282}, 24 (1992).

\bibitem{kaufmann}
	W.B. Kaufmann, P.B. Siegel, and W.R. Gibbs, Arizona State
University preprint, January 1992.

\bibitem{spain93}
	J. Nieves, E. Oset and C. Garcia-Recio,
	Nucl. Phys. {\bf  A554}, 509 (1993).

\bibitem{tabakin}
	Y.R. Kwon and F. Tabakin, Phys. Rev. C {\bf 18}, 932 (1978);
D.P. Heddle, Y.R. Kwon, and F. Tabakin, Comp. Phys. Comm. {\bf 38}, 71 (1985).

\bibitem{lpott}
	R.H. Landau, S.C. Phatak, and F. Tabakin, Ann. Phys. {\bf 78},
	299 (1973); R.H. Landau, Comp. Phys. Comm. {\bf 28} 109 (1982).

\bibitem{l+t}
	A. W. Thomas and R. H. Landau, Phys. Rep. {\bf 58},No.  3, 121
(1980); R.H. Landau and A. W. Thomas, Nucl. Phys. A{\bf 302}(1978)461.

\bibitem{coester}
	L.L. Foldy and R.A. Krajcik, Phys. Rev/ Lett. {\bf 32}, 1025
	(1974); F. Coester, Helv. Phys. Acta. {\bf 38}, 7, (1965).


\bibitem{qmII}
	R.H. Landau, {\ Quantum Mechanics II}, J. Wiley, New York (1990).

\bibitem{landau83}
	R.H. Landau, Phys. Rev. C {\bf 27}, 2191 (1983).

\bibitem{mach}
	A. Ciepl\'{y}, M. Gmitro, R. Mach, S.S. Kamalov, Phys. Rev. C
{\bf 44}, 713 (1991); M. Gmitro, S.S. Kamalov, and R. Mach, Phys. Rev.
C {\bf 36}, 1105 (1987).

\bibitem{gal}
	G. Kalbermann, E. Friedman, A Gal, and C.J. Batty, Nucl. Phys.
{\bf A503}, 632 (1989); C.J. Batty, E. Friedman, A Gal, and G.
Kalbermann, Nucl. Phys. {\bf A535}, 548 (1991).


\bibitem{vac}
	J. Blomqvist, Nucl. Phys., B{\bf 48} 95 (1972).

\bibitem{eli}
	E. Friedman, personal communication.

\bibitem{rhos}
	H. De Vries, C. W. De Jager, and C. De Vries,
	{\em Atomic Data and Nuclear Data Tables} {\bf 36}, 495
	(1987).

\bibitem{wfp}
	E. Hern\`{a}ndez and A. Mondragon, Phys. Rev. C {\bf 29}, 722 (1984).

\bibitem{bethe}
	H.A. Bethe and R. Jackiw, {\em Intermediate Quantum
Mechanics}, W.A. Benjamin, Reading (1968).

\bibitem{seki}
	R. Seki, K. Masutani, M. Oka, and K. Yazaki, Phys. Lett. {\bf
97B,} 200 (1980); R. Seki and K. Masutani, Phys. Rev. C {\bf 27}, 2799
(1983); R. Seki, K. Masutani, and K. Yazaki,  Phys. Rev. C {\bf 27}, 2817
(1983).

\bibitem{1sO16}
      I. Schwanner, R. Abela, G. Backenstoss, W. Kowald, P.
Pavlopoulos, L. Tauscher, H.J. Weyer, P. Bl\"{u}m, M. D\"{o}rr, W.
Fetscher, D. Gotta, R. Guigas, H. Koch, H. Poth, G. Schmidt, and H.
Ullrich, Phys. Lett. {\bf 96B}, 268 (1980).

\bibitem{2pO16}
      G.De Chambrier, W. Beer, F.W.N. De Boer, K. Bos, A.I. Egorov, M.
Eckhause, K.L. Giovanetti, P.F.A. Goudsmit, B. Jeckelmann, K.E.
Kir'yanov, L.N. Kodurova, L. Lapina, H.J. Leisi, V.I. Marushenko, A.F.
Mezentsev, A.A. Petrunin, A.G. Sergeev, A.I. Smirnov, G. Strassner,
V.M. Suvorov, A. Vacchi, and D. Wieser, Nucl. Phys. {\bf A442}, 637
(1985).

\bibitem{ca40-1}
      R.J. Powers, K.C. Wang, M.V. Hoehn, E.B. Shera, H.D. Wohlfahrt,
and A.R. Kunselman, Nucl. Phys. {\bf A336}, 475 (1980).

\bibitem{ca40-2}
      C.J. Batty, S.F. Biagi, E. Friedman, S.D. Hoath, J.D. Davies.
G.J. Pyle, G.T.A. Squier, D.M. Asbury, and H. Guberman, Nucl. Phys.
{\bf A322}, 445 (1979).



\bibitem{ca44}
      C.J. Batty, S.F. Biagi, E. Friedman, S.D. Hoath, J.D. Davies,
G.J. Pyle, G.T.A. Squire, D.M. Asbury, and M. Leon, Phys. Lett. {\bf
81B}, 165 (1979).

\bibitem{bi}
      E. Friedman, H.J. Gils, H. Rebel, and Z. Majka, Phys. Rev. Lett.
{\bf 41}, 1220 (1978).

\bibitem{konijn}
	J. Konijn, C.T.A.M. de Laat, A. Taal, and J.H. Koch , Nucl.
Phys. {\bf A519}, 773 (1990).

\bibitem{prob} Because we use the Gamow state normalization
(\ref{norm})-(\ref{renorm}), it can be argued that the proper
probability density is $(\mbox{Re~}\psi_{l}(k))^{2}
- (\mbox{Im~}\psi_{l}(k))^{2}$. The difference is not significant for the
present case.

%\bibitem{barr} B.R. Barrett and M.W. Kirson, in
%{\it Advances in Nuclear Physics},
%    eds. M. Baranger and E. Vogt (Plenum Press, New York, 1973), Vol. 6,
%%p.219.
\end{references}
\end{document}